# Not Even Wrong: On the Limits of Prediction as Explanation in Cognitive Science

[1][†]Orr, M., [1]Cranford, D., [1]Ford, K., [1]Gluck, K., [1]Hancock, W., [2]Lebiere, C. [1]Pirolli, P., [3]Ritter, F., [4]Stocco, A.

[1]Florida Institute for Human and Machine Cognition
[2]Carnegie Mellon University
[3]Pennsylvania State University
[4]University of Washington
[†]morr@ihmc.org

**Abstract**
We offer a comment on the Centaur (Binz et al., 2025) transformer-based model of human behavior. In particular, Centaur was cast as a path towards unified theories of cognition. We offer a counter claim with supporting argument: Centaur is a path divergent from unified theories of cognition, one that moves towards a unified model of behavior sans cognition.

**Introduction**
This article is a comment on Centaur (Binz et al., 2025), a novel model of, seemingly, human cognition. Its novelty comes from its ability to predict human behavior as well as its nature: Centaur is a pre-trained transformer artificial neural network with additional fine tuning specific for data on human behavior that was generated via experimental tasks. Centaur is presented as a "foundation model" of human cognition: when trained with data that captured over 10 million single trials of data from approximately 60 thousand participants across 160 typical cognitive tasks, Centaur was able to outperform every single, task-specific, state-of-the-art computational model[1]. Because of this, Centaur was originally claimed (Binz et al., 2024), by its authors, to be the first true unified model of human cognition, a claim that was tempered upon publication (Binz et al., 2025)[2]. Although we acknowledge the novelty and impressive scale of the approach, in this commentary we join the growing chorus of responses from the cognitive science community (Bowers et al., 2025; O'Grady, 2025) regarding Centaur's deficiencies, limitations, omissions, and contentious claims.

We present various, overlapping perspectives on Centaur, all of which point to deficiencies, limitations, or contentions not yet noted. Throughout, we may use Centaur or CP to mean the paper, the model, the claims and, really, anything Centaur related.

---

[1] A translation of the experimental paradigm and its behavioral data to natural language is a requirement for any experiment simulated by Centaur.
[2] The original article was in Psych Arxiv but in the current Nature paper, they only use the term "unified model of cognition" in the conclusion and it is not stated as a strong claim nor is the term well defined.

**Contributions**

First, we would like to highlight a number of contributions that we all find unquestionably positive and noteworthy.

*Collaborative Computational Science:* Centaur is a large, cooperative effort that produced a functioning computational system. The Centaur effort, then, may add to the lessons learned on how to organize scientists and science in cognitive, neural and psychological sciences. There are various kinds of efforts in this general space, e.g., the human connectome, the cognitive atlas. Centaur aligns closest, by self-assignment, to large-scale community efforts for modeling human cognition with cognitive architectures, e.g., Soar and ACT-R.

*Large-Scale, Harmonized HSR Data Repository:* Centaur co-evolved with the Psych-101 data set (160 experiments, 10 million single trial responses by ~60 K participants, https://huggingface.co/datasets/marcelbinz/Psych-101). Thus Centaur, in conjunction with Psych-101, contributes to the arena of large, unified datasets of human behavior. An impressive feature of Centaur/Psych 101 is the harmonization of data across experiments.

*Ease of Use:* Centaur affords ease of use in a particular way. It shifts the design decisions from the human cognitive modeler/designer to the Centaur methodology requirements: (i) a Pre-trained LLM, (ii) a way of translating an experimental paradigm and its human performance data into natural language, (iii) a method for fine-tuning the pre-trained LLM on the natural language data. This methodological approach removes the burden of model design and implementation from the human cognitive modeler/designer. In fact, the Centaur method removes the need for a cognitive modeler of any kind

*Automated Scientific Practice:* Centaur in conjunction with the Psych-101 dataset adds another approach to automated science practices and methods (Musslick, et al, 2023) in cognitive science and psychology (see Rmus, et al. 2025 and Dillion et al. 2023 for other recent approaches).

In sum, Centaur makes valuable contributions. Yet, as expected with any given novel approach in a discipline, such contributions rarely come without issues. We detail some of these next.

**Concerns**

*A Category Error:* Centaur claims to be both a domain-general computational *model* of cognition and a unified *model* of cognition. It is claimed to be *a* (not *the*, but *a*) next step towards a unified **theory** of cognition.

The logic of the Centaur claim is straight-forward: (i) most ML and cognitive science models are domain specific (e.g., designed to be good at one thing); (ii) to understand human minds, we must move towards an integrated theory; (iii) a key step in the direction of a unified theory of cognition is the building of a "computational model that can predict and simulate human behaviour in any domain" (Centaur, pg. 1, pr. 3).

At face value, we agree with this logic. But the details matter. Upon serious consideration of Centaur, we are compelled to put forth a counter claim: Centaur is a path divergent from unified theories of cognition, one that moves towards a unified **model of behavior** *sans* cognition.

To understand our claim, it helps to study the history of Centaur. Centaur was developed in the spirit of what is called integrative benchmarking in computational neuroscience (Schrimpf, 2020)[3]. Integrative benchmarking has three key components: (i) a defined domain of intelligence (e.g., vision in primates), (ii) a set of isolated experimental behavioral and neural data on sub-domains (e.g., color vision, motion, object recognition, etc.), (iii) a set of isolated, neurally mechanistic models that map to the sub-domains. By this framing, an integrative/unified model is any model that is neurally mechanistic and can capture regularities across the set of isolated experimental data across all sub-domains. A sufficient benchmark, notably, is the set of isolated experimental data in existence at any point in time. Integrative benchmarking reflects an iterative process that is designed to systematically and steadily grow better integrated/unified models.

The parallel to cognitive science is exact if one allows the domain to be human cognition. As Newell noted more than 50 years ago (Newell, 1973), research in human cognition was replete with isolated data and isolated mechanistic models (or mathematical models approximating mechanisms) with little in the way of integration or unification. His proposed solution was the construction of unified theories of cognition, any one of which should be able to explain, in mechanistic, cognitive terms, some of the data patterns found across the set of isolated data. Iteration has happened; we now have a set of mature cognitive architectures, with Soar and ACT-R as the most prominent. Newell's criteria (the Newell Test) were born from these concerns.

Yet, the parallel to Centaur is fraught. Centaur is a non-mechanistic model with respect to the domain of human cognition in that it does not propose the mechanisms of cognition. Ipso facto, we have a mismatch between integrative benchmarking, a la neuroscience or a la Newell, and Centaur. It is a category error to claim a parallel among integrative benchmarking and Centaur[4]. By itself, this error is not catastrophic for Centaur the model—e.g., it might be brushed off by saying that Centaur is a very good model, in fact; it predicts human behavior and without cognitive mechanisms to boot.

---

[3] Schrimpf was referenced in the Centaur paper.

[4] The authors of the Centaur paper do not make this parallel explicit; we have asserted it from our reading of the original preprint (Binz, et al. 2024) and final published version (Binz, et al. 2025) of the Centaur paper in conjunction with Schrimpf (2020).

The real damage, however, comes from the brute force of the argument that *mechanism* is the bedrock of theory in cognition. The lack of explicit mechanism in Centaur effectively precludes any claims that there exists a path from Centaur, as it is today, to a unified theory of cognition. Without mechanisms that can be expressed in cognitive terms, Centaur's ultimate goal of being a unified *theory* of cognition is, by definition, unattainable .

Notice that Centaur could potentially model *any* data from *any* domain of behavior, not just the cognitive domain and not just human-generated data. The same cannot be said of the models generated from existing unified theories of cognition. A unified theory of cognition, e.g., Soar or ACT-R, by its very nature, excludes the modeling of some data. No serious existing theory of cognition, for example, allows episodic memories that become more available as time passes, or the reinforcement of actions that do not result in rewards, or better memory for longer lists of words compared to shorter lists of words, and so on. Although these violations of human behavior may seem canned, trivial examples, their importance is apparent when one acknowledges that none of them would be impossible to reproduce for Centaur[5]. In fact, Centaur will likely model well any data that encode some degree of statistical regularity. Wither the cognitive?

Let's circle back to our original counter claim: Centaur is on a path towards a unified model of behavior *sans* cognition, not towards a unified model of cognition. If our claim is correct, then, what is a unified model of behavior *sans* cognition? Whatever is meant by unification? And, why is it a laudable scientific objective?

It is true that other disciplines and historical foci in the behavioral sciences, in the right light, seem similar to Centaurian unification (e.g., American behaviorism, behavioral economics and decision-theory, social epidemiology—Simon (1969) dubbed such efforts as explanation from the outside). But, across these efforts, unification isn't the putative objective; predicting, describing or explaining behavior *sans* cognition from an agent's environment alone (plus a goal) is the objective. Contrast this to the neuroscience/Newell approach—combining, integrating, or synthesizing cognitive or neural mechanisms in the service of better dynamic process models of neural and behavioral data. The criteria for neuroscience/Newell unification are twofold: (i) have neural or cognitive mechanism(s) that one can unify over, and (ii) reproduce regularities in human behavior.

---

[5] See Bowers et al., 2025 for tests of these kinds of examples.

Centaur Response for arXiv

For Centaur, unification may just mean to reproduce human behavioral data across experiments with a single statistical model. Other terms may be more fitting—compression, reduction, compilation—because behavior alone is not amenable to unification. And, statistical modeling does not imply the unification of behavioral data. A statistical model that captures well the statistics across independent experiments would point to the potential for sameness of process not unification across different processes, ceteris paribus. Yet, the value of Centaur is supposed to be its ability to unify across different kinds of experiments, and, by implication, different cognitive processes and mechanisms.

The flip-side of unification is generalization, or the hope of generalization. Newellian unification, by example, hopes that unified theories of human cognition will afford, over time, more general models of human cognition and models that exhibit generality across many tasks. Taken at face value, Centaur shines in terms of its generality—one model that performs well on 160 experimental tasks. But, in a sense, this is an illusion. Centaur, by definition, performs only one task—an estimate of the most likely next token. The generality of Centaur comes from the method for transforming experimental paradigms and associated behavioral data into natural language data, and letting the machinery of transformers do the rest.

In short, Centaur is one task to unify them all. Perhaps this is the kind of unification envisioned by the authors of Centaur?

*Newell's Criteria or Newell's Test:* The Newell Test (Newell, 1990; Anderson & Lebiere, 2003) is invoked in the Centaur paper as part of their claim that Centaur is a model of cognition. A strong claim by the authors of Centaur, in the Supplementary Materials, is that Centaur is the first model to satisfy the majority of these criteria. Below is a brief discussion of our evaluation of Centaur on the Newell Test criteria rebutting this claim, followed by a tabular listing (Table 1) that differentiates between Centaur as a fixed system vs the underlying transformer model.

1. *Behave as an almost arbitrary function of the environment*: This criterion is, ultimately, about learning and adaptability. It is true that Centaur learns to mimic human behavior in a variety of experimental settings. However, the authors miss the ***fundamental distinction between the process that created the system* (i.e., the training of Centaur) *and the system itself*** (the resulting frozen system: Binz et al., 2024)**.** Given enough parameters and training time, multi-layer transformers can be trained to be an almost arbitrary function of their inputs but in the absence of real time learning the resulting system is frozen. This distinction is applicable to most Newell Test criteria. The Table 1 (below) provides a brief comparison of the claims that can be made from either perspective.

In this model, there are 70 billion (16 bit) parameters (Llama Team, 2024). The data is 10 million choices (psych-101 cite), with what appears to be no more than 32 choices (5 bits max on average). The parameters start at 7000x the number of data points, but when you account for how much bigger the parameters are, there is 14M x more model than data.

2. *Operate in real time.* Again, Newell's claim is about adaptability to the environment. The interpretation of "real time" is meaningless, given boundless computational power and the lack of commitment to theoretical constraints on a time course of processing *in simulated time*. Centaur is either fast or slow, depending on the hardware it is, as it does not commit on how long a specific process (e.g., inhibiting a response) should last. Note that this constraint also applies to learning processes—all of their responses take the same time, which humans do not.

3. *Exhibit rational, that is, effective adaptive behavior.* As noted in point #1 above, once trained the resulting system is largely incapable of even the most straightforward adaptation to an additional task. To some extent reinforcement learning from human feedback used for fine tuning the pre-trained transformers partially addresses this issue but in a way that is fundamentally implausible, e.g., by fine tuning Centaur on a large share of the human experimental data, something that is not available to actual human participants.

4. *Use vast amounts of knowledge about the environment.* Centaur has assimilated vast amounts of data about the conditions presented to participants in the 160 retrospective datasets and the more than 10 million responses made by the participants in those studies. However, the system doesn't learn really from the environment, as a real, situated agent would, but instead depends on batch training. This point also raises epistemological questions. Does Centaur deal with "knowledge" or data or information?

5. *Behave robustly in the face of error, the unexpected, and the unknown.* Considering the extreme comprehensiveness of its training, it is unclear the extent to which Centaur generalizes to the unexpected and the unknown. Transformers often have been proven to be unable to adjust to error and use even the most simplistic feedback.

6. *Integrate diverse knowledge*. All data input into Centaur sequences of language tokens, although diverse enough to seemingly encompass much human knowledge, and multi-modal generative AI systems also exist. Newell originally meant this to refer to symbolic capacity for knowledge combination. The extent of this capability in transformers and the internal structure of this knowledge is still unclear. And, again, as with point #4 above, we could raise epistemological questions. Does Centaur deal with knowledge or data or information?

7. *Use (natural) language.* Centaur certainly does as far as syntax goes, although other aspects such as semantics are less clear, as they are derived solely from the statistics of language (at least in pure LLMs such as this one) without external grounding, which sometimes results in clear inconsistencies.
8. *Exhibit self-awareness and a sense of self.* While the model can profess a sense of awareness, there is no evidence that it is any deeper than reflecting its training data. In particular, it is incapable of metacognition, i.e., true reflection about its operations, and does not have an obvious mapping to theories of consciousness and metacognition.
9. *Learn from its environment.* As mentioned in point #1 above, while transformers are capable of almost arbitrary learning, Centaur itself, i.e., the pre-trained and fine-tuned transformer, **does not learn** in any meaningful way. In particular, it doesn't exhibit the various kinds of human learning exhibited in embodied domains such as interactive task learning (Gluck & Laird, 2019), including learning of semantic and episodic memories, acquisition of procedural skills, joint co-learning and co-teaching, and phenomena such as priming, step skipping, and conditioning.
10. *Acquire capabilities through development.* No claims are made by Centaur's creators about this criterion although the development of representations during transformer training could potentially be mapped to the development of capabilities during the early years of human life
11. *Arise through evolution.* No claims are made by Centaur's creators about this criterion. Given the relatively homogenous transformer architecture, there is no direct equivalent to the gradual development of specialized neural capabilities in biological evolution.
12. *Be realizable within the brain.* We find that claiming that "Centaur still represents the current state-of-the-art when looking at neural alignment to human subjects" is serious overreach, considering the speculative nature of the match between human neural data and the internal activity of Centaur's neurons. Most importantly, no claim was made to understand the mapping from Centaur to the human brain at a system (or architectural) level, something that has been systematically tested for some cognitive architectures and the Common Model of Cognition (e.g., Hake et al., 2022; Sibert et al., 2022; Stocco et al. 2021).

TABLE 1: *Newell's Criteria for Theories of Mind, differentially applied to either the system that underlies Centaur (a specific transformer) and the final product (a unified model of cognition).*

| Criterion | Centaur-the-system (LLAMA architecture) | Centaur-the-product (the abilities of LLAMA with large psych data) |
|---|---|---|
| 1 *Behave as an almost arbitrary function of the environment*: | Yes. It can approximate any function. It can be used to respond to new information and can be fine-tuned with RL. | No. Knowledge is crystallized and can only be added in the form of expanded inputs. |
| 2 *Operate in real time* | N/A the underlying architecture is a computational framework, its speed depends on the underlying hardware. | Unspecified. No obvious mapping between input processes and cognitive time. |
| 3 *Exhibit rational, that is, effective adaptive behavior* | N/A. The LLAMA transformer can be trained to produce irrational behavior; it is an unconstrained input-output sequence mapping. | Unspecified. For instance, "bounded rationality" is meaningful only if the bounds (i.e., cognitive costs) are specified. |
| 4 *Use vast amounts of knowledge about the environment* | Yes. (Caveat: may substitute "data", "information", or other similar terms for "knowledge.") | Yes. (Caveat: may substitute "data", "information", or other similar terms for "knowledge.") |
| 5 *Behave robustly in the face of error, the unexpected, and the unknown* | Yes. | Yes (but hard to judge, since we do not have any evidence that is independent of the training). |
| 6 *Integrate diverse knowledge* | Maybe. Possibly not diverse enough; unlikely that certain aspects of cognition are recovered from text alone, e.g. spatial cognition seems to exist independent of language abilities. | Maybe (see the table cell directly to the left). |
| 7 *Use (natural) language* | Yes. | Yes. |
| 8 *Exhibit self-awareness and a sense of self* | No. This is an intrinsic limitation of the architecture. | No, because that would depend on Centaur-the-system (which has none). |
| 9 *Learn from its environment* | Yes; the system is designed the amass large quantities of data/information, at least in batch mode. | Not in the canonical sense of "learning", as the product is frozen. |
| 10 *Acquire capabilities through development* | Unknown. | Unknown. |
| 11 *Arise through evolution* | Unknown. | Unknown. |
| 12 *Be realizable within the brain* | Unknown. | No. |

*Measurement and its Discontents:* Centaur claims that it is a computational model that can *simulate* human behavior under the condition that there is a suitable procedure for expressing an original experiment designed for humans—its procedure and generated data—into natural language. Such translation is no small feat, yet Centaur has claimed to have found such a procedure for at least 160 experiments in cognitive psychology and related fields. We won't quibble over the procedure (it was full human translation) because we want to focus on another central point.

Let's assume it returns a correct natural language translation of an experimental procedure and its generated data. Then, for any Centaur experiment we can define *two versions*: (i) one that can be administered to both humans and Centaur (the natural language expressed version) and (ii) one that can be administered only to humans (the original that might include visual stimuli and external and self-paced timing of stimuli). To date, Centaur has yet to provide the natural language-only version to humans, and thus, has failed to provide a direct version-to-version comparison in human subjects.

We can readily imagine how it might work using the Go/No-Go task, which in Centaur's Psych-101 data includes 463 participants generating approximately 150 K trials. The human experimental data (the original) already exists. The question is, how might we administer the natural language version (the version Centaur actually processes) to human subjects? The simplest way would be to just give human participants exactly what it was that Centaur took as input during training and test. Thus, the paradigm would provide the human subjects with 90% of the Centaur version data for training with the remaining 10% for testing. That is, provide the human subjects with 135 K choices to process for training prior to testing. Here is a sample of the training phase of such a paradigm (from p 27 Supplementary Information in Binz et al., 2025):

```
"In this task, you need to emit responses to certain stimuli and omit
responses to others. You will see one of two colours, colour1 or colour2, on the
screen in each trial. You need to press button X when you see colour1 and press
nothing when you see colour2. You need to respond as quickly as possible. You
will be doing 10 practice trials followed by 350 test trials.
   [train trial 0] You see colour1 and press nothing.
   [train trial 1] You see colour2 and press <<X>> in 753.0ms.
   [train trial 2] You see colour2 and press <<X>> in 381.0ms.
   [continue for approximately 135 K trials]."
```

Here is a sample of the testing phase:

```
"In this task, you need to emit responses to certain stimuli and omit
responses to others. You will see one of two colours, colour1 or colour2, on the
```

9
Centaur Response for arXiv

```
screen in each trial. You need to press button X when you see colour1 and press
nothing when you see colour2. You need to respond as quickly as possible. You
will be doing 10 practice trials followed by 350 test trials.
    [test trial 0] You see colour1 and press ??? in ??? ms.
    [test trial 1] You see colour2 and press ??? in ??? ms.
    [test trial 2] You see colour2 and press ??? in ??? ms.
    [continue for approximately 15 K trials]."
```

The question marks denote where a response is required of the subject in the test phase of the experiment. Also, notice that in the training phase, we would have to fudge a little bit by telling the subjects that they should forgo responding because the response will be printed. We might even tell them that they should just read these as preparation for the test and that the test will require them to respond.

So, what is the point of this imagining? Why is it important? Centaur has claimed that it simulates how humans do the task. If that is true, shouldn't humans exhibit similar patterns of behavior on both versions of any Psych-101 experiment? Do humans bring to bear the same cognitive processes in both versions of the task? We have no experimental data to settle this point, but it is worth serious consideration.

Notice that Centaur's Psych-101 data, for some experiments, does include response times at the trial level. It is likely not contentious to say that these response times are not to be taken literally (as if Centaur took that many simulated milliseconds for cognitive processing). Any response time data given by Centaur is the next predicted token, full stop! (No pun intended.)

*Centaur is Non-Mechanistic and Atheoretic:* Physicist Richard Feynman discussed the subtleties of scientific theories and understanding.[6] Part of his lesson was that there is more to theoretical understanding than mere prediction. Feynman explained that two theories can be equivalent in their predictions, yet be "psychologically distinct", offering different frameworks to scientists for inspiring new ideas. Fundamental concepts in a theory may drastically change in response to even small discrepancies with data—such as the shift from a Newtonian to Einsteinian view of the universe in response to minor anomalies in Mercury's predicted orbit. The deeper "philosophies" or "understandings" of theories provide tricky ways to compute consequences quickly and generate new hypotheses.

Feynman implied that theories or models impact the psychology of doing science beyond mere prediction, and they can accelerate or stifle discovery. One paradigm of psychological and historical studies conceptualizes the broader process of scientific discovery (Klahr & Simon, 1999) as complex collaborative problem solving in which prediction plays only a part. In this paradigm, scientific discovery can be viewed as problem space search in multiple interrelated problem spaces. In its original dual-space formulation (Klahr & Dunbar, 1988; Simon, 1992), discovery involves navigating a problem space of hypotheses and a problem space of experiments. In the hypothesis space, hypotheses or laws are formulated, induced from observation, evaluated, and selected for testing. The experiment space involves designing, executing, and interpreting experiments. Making predictions and comparing them to data is just a part of the work in the experiment space. This approach has been expanded to multiple space formulations (Schunn & Klahr, 2000) including an experimental paradigm space and a representation space in which mental models such as Bohr's planetary view of atomic structure, Faraday's search for a good representation of magnetic induction of electricity, and Feynman's creation of his eponymous diagrams play crucial roles. Whether or not one accepts this particular theory (and it is likely wrong in complexity and detail), it emphasizes that scientific discovery is much more than prediction and bolsters Feynman's view that different mental models (representations) can radically advance discovery.

---

[6] Cf. https://youtu.be/NM-zWTU7X-k?si=CMUiXB6Rba04EQwq

Newell's "old-fashioned" view of science—and one we think most cognitive scientists would agree with—is that theories provide answers to certain kinds of questions including questions about predictions, explanations, and a foundation for prescriptions for desired interventions, designs, or control. Prediction, empirical coverage, accuracy refinement, and cumulation are certainly important. However, predictive accuracy and breadth of predictive coverage is not everything. Explanation (or understanding), or control of something in the world, can frequently be achieved with models that are only approximations (often deliberately abstracted). Good theories change the way we think the world works, in ways that open it up for breakthroughs in technique or technology. Just as important, theories drive the science itself, generating new questions and probes of the world. Working scientists know that data themselves are not the "ground truth" of reality, but the responses to questions, instruments, and measurements posed to a universe that always begs more questions. Theory drives the generation of innovations that create new instrumentation and collection of novel signals that allow us now to "see" black holes, gravity waves, or the processing of mental images and maps in the brain.

Currently, multilevel mechanistic approaches are common in cognitive science. Simon (1969) argued that evolutionary forces shape systems that tend to be hierarchical and nearly decomposable. Newell (1990) framed human intelligence as resulting from hierarchical system levels operating at different time scales. Contemporary philosophy of science in biology and cognition (Delgado, 2022) focuses on explanations in terms of organized operations of mechanistic parts, possibly hierarchically organized. For a given phenomenon or X→Y relationship, mechanisms "get inside the arrow linking X and Y that, if treated as a black box, might be concealing what is actually interesting or misleading about the distinguishable pattern of cooccurrence" (Delgado, 2022). Mechanistic explanations identify the organization of information-processing parts and operations that give rise to phenomena in each environmental context.

Multilevel methodological approaches include Marr's (1982) influential three levels of analysis of intelligent information processing systems (computational, algorithmic, and implementation), Newell's (1990) analysis of knowledge level systems versus symbol level systems, Dennett's intentional systems (Dennett, 1971), the distinction between rational (e.g., Bayesian) models and process models (e.g., Griffiths, et al., 2024 compared to Anderson, 1990). Different kinds of questions are asked at different methodological levels. These may include "why" questions about function, adaptation, and rationality with respect to the environment, "what" questions about information processing mechanisms, making commitments to representation, content, and processes, "how" questions that address the implementation of those information processing mechanisms in physical systems. The mapping between physical (e.g., brain or silicon) explanations and informational ones is not unique: Theories of natural selection distinguish between the genes as codes and their underlying DNA implementation. More generally, multiple levels of explanation in cognitive science are consistent with multilevel methodologies that emerged in the modern synthesis in biology (Mayr, 1983; Tinbergen, 1963).

In current practice, cognitive scientists tend to work at one methodological or mechanistic level, or on a particular slice of phenomena; unification, however, has been and remains a goal in the field of cognitive science since, at least, Newell's (1973) seminal "20 Questions" paper. Perhaps it is arguable, but this approach has been productive at least in terms of uncovering new phenomena, creating new kinds of experiments and formulating new kinds of approaches and models. For example, a comparison of psychological phenomena studied using information processing mechanistic models highlighted in 1972 by Newell (1973) appears to be completely different from the phenomena addressed by a recent monograph on rational (Bayesian) models (Griffiths et al. 2024). As Feynman lectured, theories and frameworks are more than prediction machines, they foster different kinds of understanding of the universe and new ways of testing and controlling it. Theory not only accounts for past data but accelerates the discovery of novel and previously unexpected facts, explanations, and insights. This perspective raises the question of all research methodologies, including approaches such as Centaur: How will they aid us in moving forward with theory that is progressive, productive, and improves our understanding of cognition?

As Feynman also lectured, science can progress—sometimes in big ways—when models do *not* fit the data, when they are wrong. From this reckoning, we can and should question what we would conclude if Centaur's predictions turned out poor, say only at chance. What would this falsify, for example? The training data? The weight optimization scheme(s)? The fine-tuning method? The transformer architecture itself? Any of the 70 billion parameters of the underlying transformer? How would we address such falsification in a principled way, one that is intimately tied to some theoretical model of cognition? Would we have such insights into Centaur's failure that leads us to a better understanding of cognition? And, given transformers' capacity to regurgitate statistical regularities in the data, could they even make substantially incorrect predictions? We think such considerations (or musings) lead us to an important fact[7]: *Centaur isn't even wrong!*

*Claims of neural alignment are overblown:* Their claim that Centaur's internal representations become more aligned to human neural activity through fine-tuning is a stretch, at best. Internal representations were defined by extracting the hidden-layer vectors for each transformer layer in Cenatur/LLAMA, and then reducing the dimensions via principal components analysis. These internal representations thus reflect the statistics behind producing the output. The reduced internal representations were then entered into a series of regularized linear regression models to predict the mean brain activity of different cortical parcels (from the Glasser 2016 atlas); a different model was fit for every participant and region. On average, Centaur produced better results than LLAMA. Is Centaur, therefore, better aligned to neural activity than LLAMA only because it is also a better predictor of human behavior?

It should be noted that the connection between Centaur's inner representations and human brain data is indirect and highly mediated: the same representation has to be passed through different models for each individual and region to achieve significant correlations.

What does it mean, then, to become "more aligned" to human neural activity anyway? At face value, the authors seem to mean that Centaur, the fine-tuned LLM, aligns better in comparison to LLAMA without fine-tuning. But, perhaps they are implying something deeper. Does it mean that the internal representations of Centaur somehow map to the functional anatomy of the human brain? Certainly not; the authors do not suggest any topological similarity between brain anatomy and Centaur representations. Is it that the brain regions associated with performing certain tasks are reflected within how Centaur performs a task? Unlikely; the analysis simply implies that the internal representations of the fine-tuned model contain *marginally more useful information* than the standard model to predict human brain activity.

---

[7] We co-opted the epistemological construct of something being not even wrong from Wolfgang Ernst Pauli (Peierls, 1960).

What it really measures, we think, is the functional correlation with the task, something that is necessarily reflected in both the human brain and Centaur. In other words, in Marr's terms, it reflects a computational analysis of the tasks but does not provide any additional insights into the algorithmic and implementational solutions of the human brain.

Beyond this analysis, no other evidence is provided of substantive similarity between Centaur's inner representations and human brain function. LLM internals do not align with what we know of cognition, let alone the brain. For example, LLMs do not learn in the same way as humans; they do not have, technically, a long-term episodic memory; their memory is limited to the size of their context window, and in LLAMA, the context window is immense, making it a de facto unlimited external working memory. The transformer architecture, on which all of these systems are based, is a fixed architecture that learns complex matrix operations of its large inputs. Centaur's learning and prediction abilities capitalize on this system.

In their attempt to show alignment between Centaur internal representations and human neural activity they chose a method that supports their claims and shows good results for Centaur, but which provides no explanatory power about how those representations align with neural structures. This undermines decades of research that has performed detailed experimentation and matching between brain structures and architectural mechanisms to provide explanatory power (Borst, et al., 2013; Borst, et al., 2015; Anderson, et al., 2008; Stocco, et al., 2010; Stocco, et al., 2024). Using a regression to force Centaur's internal representations into the space of neural activity provides no understanding of what the components of the representations actually reflect or how they map to brain regions. While their results may be preliminary, it currently provides no evaluation of validity. In a sense, one can wonder whether all Centaur is doing is predicting task representations as in "task descriptions".

**Conclusion**[8]

One should not be surprised by the performance of Centaur. Transformers can extract remarkable amounts of statistical regularities in complex datasets (weather forecasting, Pathak, et al. 2022; protein structures, Jumper, et al. 2021). Why shouldn't Centaur do well in predicting the next token for data that captures the statistical regularities between stimulus and response in human subjects (as captured by the natural language translation of the experimental paradigm)? Given the number of free parameters in Centaur, it makes sense that it can capture such regularities. Of course, a thorough analysis of the human experiment data—for example, a measure that captures the degree of person-task variability—would be useful for better understanding what Centaur is up against statistically. Many less parameters, say 1.5 Billion or 500 Million for example, might suffice?

---

[8] Please see our views on the positive contributions of Centaur in the Introduction. We reserve this Conclusion to summarize our concerns alone.

One should, however, be surprised by the limited scope of the kinds of cognitive models included in the Centaur comparison set. Centaur claimed to use a "collection of domain-specific models that represent the state-of-the-art in the cognitive-science literature"[9] (pg 3. pr. 1, Binz et al., 2025). Yet, there was no coverage of cognitive models based on state-of-the-art cognitive architectures for those same tasks. Nor was there coverage of higher density process data (e.g., verbal protocols, eye tracking) which has been commonplace in cognitive science over the last half century. Missing was coverage of the many dozens of other more complex tasks to which various models, including those implemented in cognitive architectures, have provided detailed mechanistic, explanatory accounts of human cognitive processes. These oversights imply some degree of ignorance of the history of cognitive architectures and cognitive architecture's unique place in the quest for unified theories of cognition (see a recent review of cognitive architectures, Kotseruba and Tsotsos, (2025)).

Our key recommendation for pushing Centaur forward is to focus on refining its predictive capabilities in respect to some suitable purpose. Suitable purposes would likely be in contexts where either low criticality (in terms of safety or reliability, e.g., non-military, non-transport, non-electrical grid contexts) or that are dynamically stable and insensitivity to extrapolation issues. Moving forward in this way would free Centaur from nonsensical theoretical obligations (about cognition) and allow it to home in on where it has potential to contribute useful applications in human behavior (e.g., personalization of prediction for greater intervention and prevention efforts in policy domains such as disaster response, medicine, psychology).

We close with a cautionary tale on the game of ontological chairs. Some claim that large language models, such as the basis for Centaur, have consciousness. Why? Because LLMs may describe their feelings when asked. This is a fallacy, clearly. A portion of training data for LLMs, including Centaur, comes from literature and socially directed internet content (social media, self-help, newspapers, blogs, etc.). So, naturally any allusion to conscious LLMs is an illusion; it's a deep fake on steroids[10]. We have a useful parallel to Centaur. Its training method, particularly the fine-tuning approach, used data from approximately 10 million experimental responses from 160 experiments that included approximately 60 thousand participants; training used 90% of the data (that is 10% holdout). Imagine watching Centaur in action; it would seem as if it was really considering its responses, etc., maybe even as if it were a human. However, just as with consciousness, this is an illusion, but an illusion that may entice even the most careful researchers into fantasy.

---

[9] The models were titled (all run in PyTorch): generalized context model, prospect theory model, hyperbolic discounting model, dual-systems model, Rescorla-Wagner model, linear regression model, weighted-additive model, decision-updated reference point-model, odd-one-out model, Multi-task reinforcement learning model, GP-UCB model, rational model, lookup table model.

[10] We stole this notion from Christof Koch (March 24, 2025, Mindscape podcast by Sean Carroll).

Perhaps the answer to all of our questions lies within Centaur itself. We might just ask it?

**References**

Anderson, J. R. (1990). *The adaptive character of thought.* Lawrence Erlbaum Associates.

Anderson, J. R., Fincham, J. M., Qin, Y., & Stocco, A. (2008). A central circuit of the mind. Trends in cognitive sciences, 12(4), 136-143.

Anderson, J. R. & Lebiere, C. The Newell test for a theory of cognition. *Behavioral and brain Sciences 26*, 587–601 (2003).

Binz, M., Akata, E., Bethge, M., Brändle, F., Callaway, F., Coda-Forno, J., Dayan, P., Demircan, C., Eckstein, M., Éltető, N., Griffiths, T. L., Haridi, S., Jagadish, A., Ji-An, L., Kipnis, A. D., Kumar, S., Ludwig, T., Mathony, M., Mattar, M. G., … Schulz, E. (2024). *Centaur: A foundation model of human cognition*. PsyArxiv https://doi.org/10.31234/osf.io/d6jeb

Binz, M., Akata, E., Bethge, M., Brändle, F., Callaway, F., Coda-Forno, J., Dayan, P., Demircan, C., Eckstein, M. K., Éltető, N., Griffiths, T. L., Haridi, S., Jagadish, A. K., Ji-An, L., Kipnis, A., Kumar, S., Ludwig, T., Mathony, M., Mattar, M., … Schulz, E. (2025). A foundation model to predict and capture human cognition. *Nature*, 1–8. https://doi.org/10.1038/s41586-025-09215-4

Binz, M., Dasgupta, I., Jagadish, A. K., Botvinick, M., Wang, J. X., & Schulz, E. (2024). Meta-learned models of cognition. *Behavioral and Brain Sciences, 47*, e147.

Borst, J. P., & Anderson, J. R. (2013). Using model-based functional MRI to locate working memory updates and declarative memory retrievals in the fronto-parietal network. Proceedings of the National Academy of Sciences, 110(5), 1628-1633.

Borst, J. P., Nijboer, M., Taatgen, N. A., van Rijn, H., & Anderson, J. R. (2015). Using data-driven model-brain mappings to constrain formal models of cognition. PLoS One, 10(3), e0119673.

Bowers, J. S., Puebla, G., Thorat, S., Tsetsos, K., & Ludwig, C. J. H. (2025, July 7). Centaur: A model without a theory. https://doi.org/10.31234/osf.io/v9w37_v3

Delgado, H. E. (2022). On the ontological status of mechanisms and processes in the social world. *Foundations of Science, 27*(3), 987-1000. https://doi.org/10.1007/s10699-021-09787-0

Dennett, D. C. (1971). Intentional systems. The Journal of Philosophy, 68(4), 87–106. https://doi.org/10.2307/2025382

Gluck, K. A., & Laird, J. E. (Eds.). (2019). Interactive task learning: Humans, robots, and agents acquiring new tasks through natural interactions. Strüngmann Forum Report, vol. 26, J. Lupp, series ed. Cambridge, MA: MIT Press. Open Access: https://esforum.de/publications/sfr26_Interactive_Task_Learning.html

Griffiths, T. L., Chater, N., & Tenenbaum, J. (2024). *Bayesian models of cognition : reverse engineering the mind*. The MIT Press.

Hake, H. S., Sibert, C., & Stocco, A. (2022). Inferring a Cognitive Architecture from Multitask Neuroimaging Data: A Data‐Driven Test of the Common Model of Cognition Using Granger Causality. Topics in Cognitive Science, 14(4), 845-859.

Jumper, J., Evans, R., Pritzel, A., Green, T., Figurnov, M., Ronneberger, O., Tunyasuvunakool, K., Bates, R., Žídek, A., Potapenko, A., Bridgland, A., Meyer, C., Kohl, S. A. A., Ballard, A. J., Cowie, A., Romera-Paredes, B., Nikolov, S., Jain, R., Adler, J., … Hassabis, D. (2021). Highly accurate protein structure prediction with AlphaFold. *Nature, 596*(7873), 583–589. https://doi.org/10.1038/s41586-021-03819-2

Klahr, D., & Dunbar, K. (1988). Dual space search during scientific reasoning. Cognitive Science, 12(1), 1-48

Klahr, D., & Simon, H. A. (1999). Studies of scientific discovery: Complementary approaches and convergent findings. Psychol Bull, 125(5), 524-543. https://doi.org/10.1037/0033-2909.125.5.524

Kotseruba, I., & Tsotsos, J. K. (in press). The computational evolution of cognitive architectures. Oxford, UK: Oxford University Press.

Llama Team. (2024). The Llama 3 herd of models: arXiv preprint arXiv:2407.21783.

Lewis, C. (2025). *Artificial Psychology: Learning from the unexpected capabilities of large language models*: Springer.

Marr, D. (1982). *Vision*. W.H. Freedman

Mayr, E. (1983). How to carry out the adaptationist program? *American Naturalist, 121,* 324-334.

Newell, A. (1973). You can't play 20 questions with nature and win: Projective comments on the paper of this symposium. In W. G. Chase (Ed.), *Visual information processing*. Academic Press

Newell, A. (1990). *Unified theories of cognition*. Harvard University Press

Newell, A., & Rosenbloom, P. S. (1981). Mechanisms of skill acquisition and the law of practice. In J. R. Anderson (Ed.), *Cognitive skills and their acquisition* (pp. 1-51). Hillsdale, NJ: Erlbaum.

O'Grady, C. (2025). Researchers claim their AI model simulates the human mind. Others are skeptical. Science (News-Technology), July 2, 2025. https://doi.org/doi:10.1126/science.z6czuhe

Peierls, R. E. (1960). Wolfgang Ernst Pauli, 1900-1958. *Biographical Memoirs of Fellows of the Royal Society, 5,* 174-192. https://doi.org/doi:10.1098/rsbm.1960.0014

Pew, R. W., & Mavor, A. S. (Eds.). (2007). *Human-system integration in the system development process: A new look*. Washington, DC: National Academy Press. books.nap.edu/catalog/11893, checked May 2019.

Pew, R. W., & Mavor, A. S. (Eds.). (1998). *Modeling human and organizational behavior: Application to military simulations*. Washington, DC: National Academy Press. books.nap.edu/catalog/6173, checked Feb 2020.

Ritter, F. E., & Schooler, L. J. (2001). The learning curve. In W. Kintch, N. Smelser & P. Baltes (Eds.), *International encyclopedia of the social and behavioral sciences* (Vol. 13, pp. 8602-8605). Amsterdam: Pergamon.

Ritter, F. E., Shadbolt, N. R., Elliman, D., Young, R. M., Gobet, F., & Baxter, G. D. (2003). *Techniques for modeling human performance in synthetic environments: A supplementary review*. Wright-Patterson Air Force Base, OH: Human Systems Information Analysis Center (HSIAC).

Schrimpf, M., Kubilius, J., Lee, M. J., Ratan Murty, N. A., Ajemian, R., & DiCarlo, J. J. (2020). Integrative Benchmarking to Advance Neurally Mechanistic Models of Human Intelligence. *Neuron*, *108*(3), 413–423. https://doi.org/10.1016/j.neuron.2020.07.040

Schunn, C., & Klahr, D. (2000). Multiple-Space Search in a more Complex Discovery Microworld

Sibert, C., Hake, H. S., & Stocco, A. (2022). The structured mind at rest: low-frequency oscillations reflect interactive dynamics between spontaneous brain activity and a common architecture for task control. Frontiers in neuroscience, 16, 832503.

Simon, H. A. (1969). *The sciences of the artificial.* MIT Press.

Simon, H. A. (1992). Scientific discovery as problem solving. International Studies in the Philosophy of Science, 6(1), 3-14. https://doi.org/10.1080/02698599208573403

Stocco, A., Lebiere, C., & Anderson, J. R. (2010). Conditional routing of information to the cortex: a model of the basal ganglia's role in cognitive coordination. Psychological review, 117(2), 541.

Stocco, A., Rice, P., Thomson, R., Smith, B., Morrison, D., & Lebiere, C. (2024). An integrated computational framework for the neurobiology of memory based on the ACT-R declarative memory system. Computational Brain & Behavior, 7(1), 129-149.

Stocco, A., Sibert, C., Steine-Hanson, Z., Koh, N., Laird, J. E., Lebiere, C. J., & Rosenbloom, P. (2021). Analysis of the human connectome data supports the notion of a "Common Model of Cognition" for human and human-like intelligence across domains. NeuroImage, 235, 118035.

Tinbergen, N. (1963). On the aims and methods of ethology. *Zeitschrift für Tierpsychologie, 20*, 410-463

Wang, S. N., & Ritter, F. E. (2023). Modeling a strategy with learning in a complex task. In *Proceedings of 21st International Conference on Cognitive Modeling (ICCM)*, 277-278.